# Premelting layer during ice growth: role of clusters

Shifan Cui[1], Haoxiang Chen[2], Zhengpu Zhao[1]

The premelting plays an important role in ice growth, but there is a significant gap in our knowledge between the atomistic premelting surface structure and the macroscopic growth mechanism. In this work, using large-scale molecular dynamics simulation, we reveal the existence of clusters on premelting surface, as an intermediate feature bridging the gap. We show the spontaneous formation and evolution of clusters, and they form a stable distribution determined by the growth rate. We demonstrate how this stable distribution is related to the growth mode of ice, connected by the growth of clusters. We come to a bilayer-by-bilayer growth mode under simulation-reachable high growth rates, but another mechanism, namely "cluster stacking", is speculated to exist at lower growth rates. Our work builds a connection between the microscopic structure of premelting layer and the macroscopic growth of ice, making a step forward to the full understanding of premelting and ice growth.

1 International Center for Quantum Materials, School of Physics, Peking University, Beijing, China.
2 School of Physics, Peking University, Beijing, China.

## Introduction

Premelting of ice, originally discovered by Faraday in 1850[1], refers to the existence of a liquid-like layer on the ice surface below its melting point. Such phenomenon plays an important role in environmental effects[2] like chemical circulation[3,4] and frost heave[5,6], and also finds its application in nanoreactors[7] and icephobic surfaces[8]. In the last decades, much work has been devoted to this liquid-like layer (usually called quasi-liquid layer, QLL) of ice, both experimentally[9-18] and theoretically[9,19-29]. Thanks to these efforts, we have gained much understanding of QLL, including its evolving temperature[9,11-13,30,31], thickness[10,20,24,25,32], and microscopic structure[20,24,26,33], especially on the static side. An excellent review is available about these developments[34].

Despite such progress in understanding static QLL, our knowledge about the dynamics of QLL during ice growth remains little. Many natural processes related to premelting (e.g. the growth of snow plates[35,36]) happen during ice growth, demanding a theoretical understanding of QLL under such conditions. Furthermore, recent experiments reported the existence of droplets on premelting surfaces during ice growth[37-39], implying new physics of QLL which is still incompletely understood. Sibley et al. simulated ice growth under premelting temperatures using a continuous model based on the equilibrium wetting theory[27], but such continuous model cannot reveal the molecular-level mechanism of ice growth, and its applicability for QLL with only 1-2 bilayer thickness[9,20,22,24,25] requires further validation. Pickering et al. studied the ice QLL during growth and evaporation using grand canonical molecular dynamics (MD) simulations[20], but failed to capture all important structural features due to small model size (as we will show).

The major challenge of understanding QLL in ice growth comes from its multi-scale nature. On the one hand, the growth of ice starts from single molecules, and QLL itself has molecular-level microstructures[20,24,30]. On the other hand, the observable features of growth (e.g. droplets) are on the scale of micrometer or larger. However, as this work will show, an intermediate-scale structural feature, namely *cluster* exists for QLL, forming a bridge between the molecular evolution of QLL and the macroscopic growth of ice.

In this work, adopting MD simulation with large models (70~280 nm width) and the highly efficient mW force-field[40], we studied the structure and evolution of QLL during ice growth under a typical premelting temperature (273.9 K, = $T_m$ - 0.7 K for mW force field[40]). Specifically, we reveal the existence of *clusters* on QLL surfaces, and observe their dynamics including spontaneous formation, growth, and vanishing. Especially, we found that the clusters from a stable distribution whose number density is determined by the growth rate. Furthermore, based on the evolutionary behavior of clusters, we propose two competing growth modes of ice, namely the bilayer-by-bilayer mode and the cluster stacking mode. We show that the competition between two modes is ultimately determined by the growth rate, where the bilayer-by-bilayer mode wins for simulation-accessible high growth rates. However, we speculate that the cluster stacking mode may win in realistic scenarios and we discuss the possibilities.

## QLL under equilibrium

Before studying QLL during ice growth, let's first have a look at the QLL under equilibrium (more accurately, not growing or evaporating). To this end, we simulated the evolution of QLL with constant molecule number, beginning from I$_h$-ice crystal with (0001) surface exposed (see Methods section). Though the evolution begins from perfect crystal is not realistic, it nevertheless provides valuable insights about QLL structure.

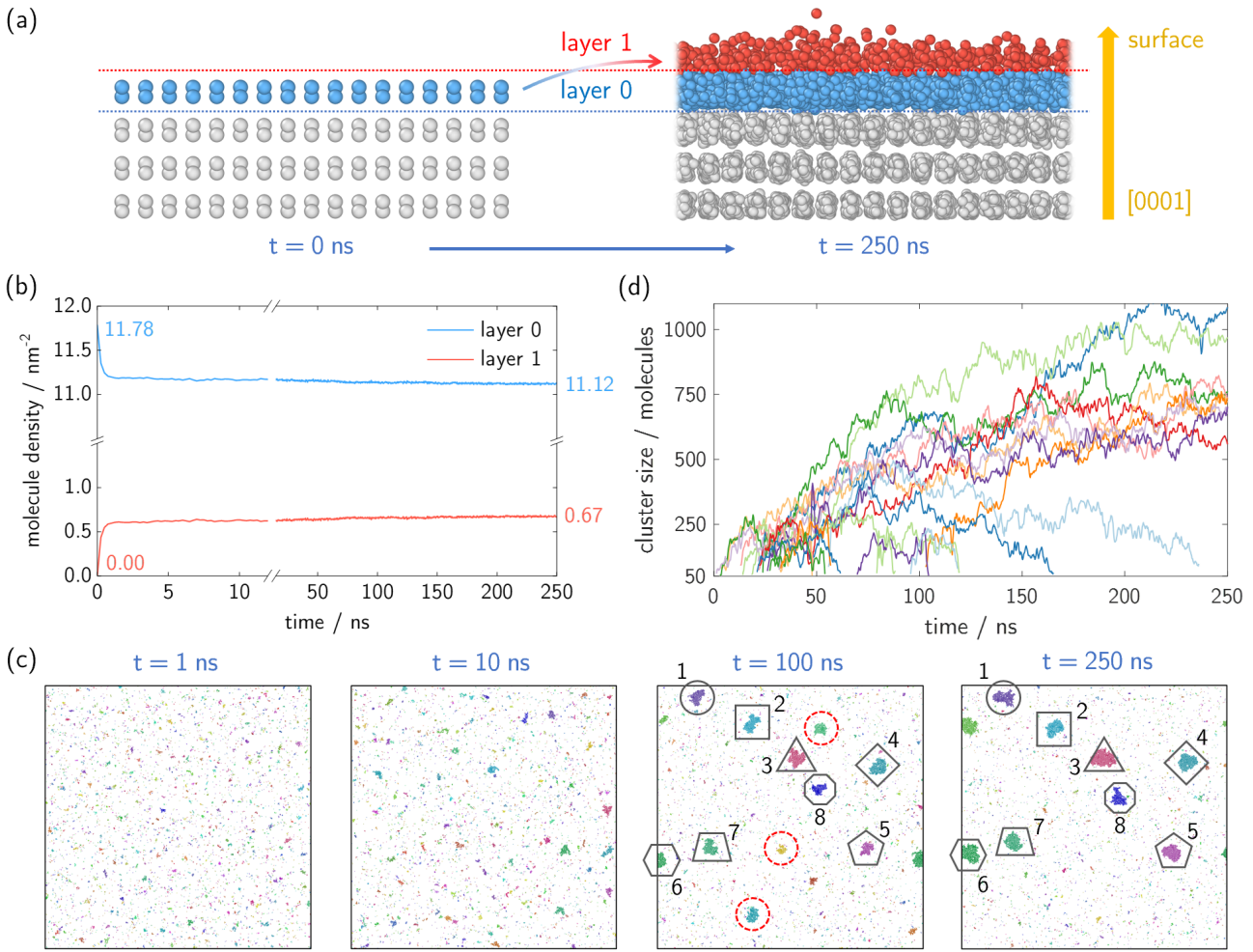

*Figure 1. Evolution of QLL under equilibrium. (a) side view of ice crystal, at the beginning (0 ns, left) and end (250 ns, right) of simulation. Each sphere represents a water molecule. Molecules in layer 0 and layer 1 are colored blue and red, respectively. Dashed lines represent the boundaries of two layers. Curved arrow indicates the direction of molecule transition (layer 0 --> layer 1). The direction of surface is shown on the right. Note that the whole model in simulation is much larger and only a part is shown here. (b) molecule number density (number of molecules per area) in two layers, respectively, as functions of time. The red and blue numbers explicitly show the values at the beginning and end of simulation. (c) horizontal distribution of molecules in layer 1, at four different times in simulation. Molecules are colored by the clusters they belong to. In the last two images some clusters are labeled with numbers and geometries for identification (see main text). Red circles in the 3$^{rd}$ image point out some clusters disappearing in the 4$^{th}$ image. The model size is ~140 x 140 nm. (d) size of clusters as functions of time; each curve represents one cluster. Clusters at different timesteps are connected by nearest-center logic (see Methods section). Only clusters with more than 50 molecules are shown. Curves are slightly smoothed for better visual appearance. Note that some colors are used repeatedly.*

Fig. 1(a) compares the structure of ice surface at beginning (left, t = 0) and end (right, t = 250 ns) of the simulation. For identifying where the molecules locate, we denote the region containing the topmost full bilayer of initial structure as "layer 0", and the region above layer 0 as "layer 1" (shown in Fig. 1(a); also see Methods section). We find that a significant number of molecules come to layer 1 during the simulation (right, red-colored particles). Fig. 1(b) further shows that layer 0 is losing molecules at a similar rate to the gaining of layer 1, indicating a spontaneous transition of molecules between two layers. Such a transition increases the volume occupied by QLL, resulting in its density being *lower* than bulk ice (which is somewhat counter-intuitive, because QLL is usually considered as something between ice and water). We find that the co-existence of molecules in two layers is thermodynamically stable, reflected by the steady ratio of molecule density between two layers (plateaus in Fig. 1(b)).

A closer look at layer 1 reveals more features in its structure. Fig. 1(c) shows how the horizontal distribution of molecules in layer 1 evolves during simulation. At the beginning (1 ns, $1^{st}$ from left), molecules in layer 1 appear randomly distributed. However, these molecules soon start to gather and form molecular "groups" (10 ns, $2^{nd}$ from left). Some of the groups continue to grow and some others shrink, resulting in fewer and larger *clusters* (100 ns, $3^{rd}$ from left). Such process proceeds, leaving a small number of large clusters (250 ns, $4^{th}$ from left). The evolution of single cluster sizes during simulation (Fig. 1(d)) further confirms our description of the whole process. The vertical distribution of molecules further shows that the clusters have well-established bilayer structures in layer 1 (see Supplementary Information Section 1), so the concepts of layers and clusters are well-behaved.

A direct conclusion of such an evolutionary process is that small clusters are unstable. Because all remaining clusters are large, small clusters (if they cannot grow to large ones) have to disappear. This is the result of thermal motion, which tends to break up clusters. This explains why large models are required for observing clusters in simulation (see Supplementary Information Section 2).

It is also noteworthy that, though single molecules can move in QLL, clusters tend to stay still once they are formed. This can be seen by comparing the last two images (100 ns and 250 ns) in Fig. 1(c): clusters with the same labels (numbers and geometries) stay at almost same locations. The physics behind here, as we believe, is that the clusters in layer 1 affect the arrangement of molecules in layer 0. Though liquid water and solid ice co-exist in both layer 0 and clusters, the region in layer 0 *covered by clusters* is almost exclusively filled with ice (see Supplementary Information Section 3). The solid-like structure under clusters provides stable support for the clusters, preventing them from moving freely. The stability of clusters plays an important role in the description of QLL, as we will see through the paper.

It should be pointed out that, however, such stability doesn't prevent clusters from disappearing. This is shown in the $3^{rd}$ image of Fig. 1(c), where some clusters (labeled with red circles) disappear later ($4^{th}$ image) in the simulation. In a thermodynamic view, we believe this reflects that a lot of small clusters are not as stable as fewer but larger ones due to higher "edge" energy (as an analogy of surface energy, proportional to the total circumference of clusters). Therefore, we can expect clusters to further merge after $4^{th}$ image if there is sufficient time. However, as we will show in the next section, this is not true for the QLL during ice growth, and such difference is what makes QLL during growth distinct in perspective of clusters.

## QLL during ice growth

Now we shall move on to the QLL during ice growth. Inspired from the last section, here we mainly focus on the clusters in QLL. To simulate ice growth, water molecules are added randomly towards the ice surface at controlled rates. In addition, some water molecules on the surface are deleted beforehand from initial structure, to guarantee no clusters form at the beginning of simulation (see Methods section). This allows us to observe the complete process of cluster formation and growth.

Fig. 2(a) illustrates the horizontal distribution of molecules in layer 1 in a typical simulation of ice growth. As we expected, no clusters are found in layer 1 at the beginning (112 ns, 1$^{st}$ from left). With water molecules continuously coming to surface, the layer 1 becomes denser and after some time clusters appear (225 ns, 2$^{nd}$ from left). Similar to the equilibrium case, a lot of clusters exist at this time but most of them are small. Clusters further merge thereafter (375 ns, 3$^{rd}$ from left), resulting in fewer and larger clusters, distributed roughly uniformly in layer 1. Clusters keep growing after that, but the number of clusters and their location do *not* change anymore (450 ns, 4$^{th}$ from left).

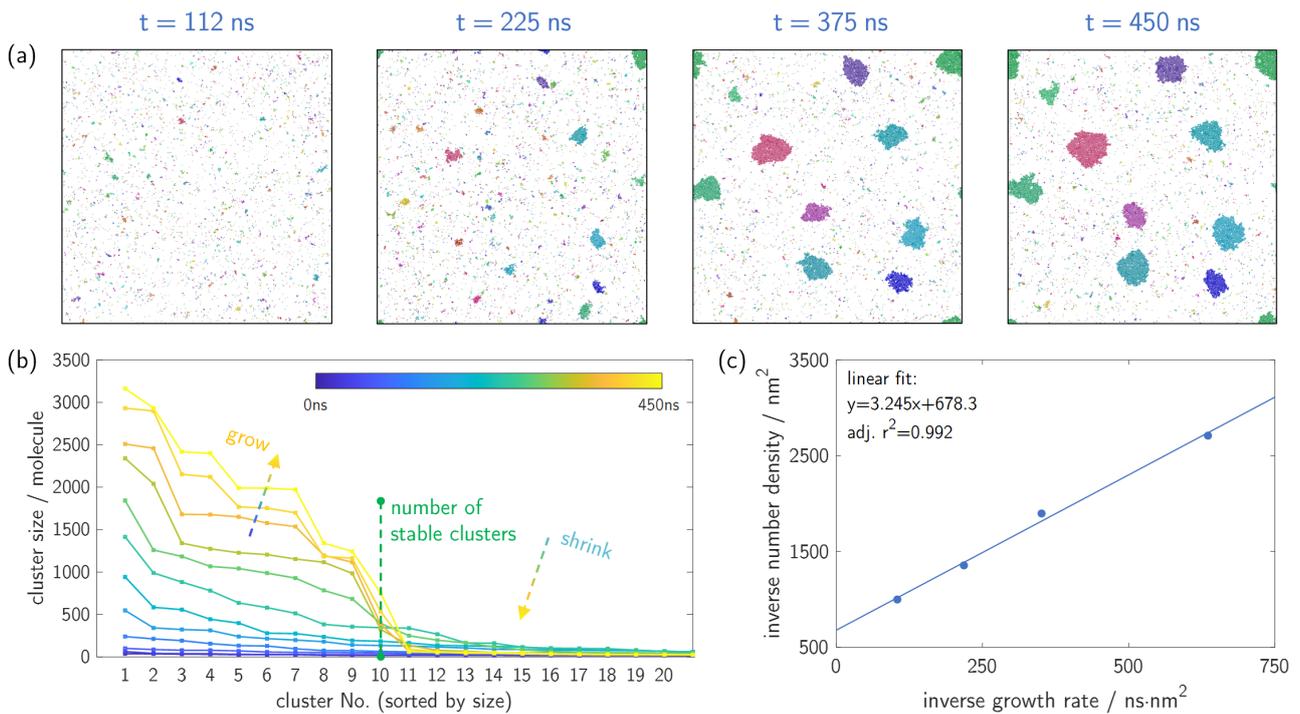

*Figure 2. Evolution and distribution of clusters during ice growth. (a) horizontal distribution of molecules in layer 1, at four different times in simulation (with inverse growth rate 351 ns·nm$^2$). Molecules are colored by the clusters they belong to. Note that the cluster distribution in 3$^{rd}$ and 4$^{th}$ images are almost the same. The model size is ~140 x 140 nm. (b) size of largest 20 clusters in the system during simulation, at uniformly spaced time points from 0 ns to 450 ns. Curves are colored by their corresponding time (see the color bar). Vertical green dashed line shows the location of "cliff" in cluster size distribution, representing the number of stable clusters. Sloping dashed arrows indicate the trend of cluster size (grow or shrink) over time. (c) relation between inverse growth rate and inverse number density of clusters. The solid line is a linear fit (result is shown at the top-left corner).*

The fact that the number and location of clusters *stop changing* after a certain time (like the 3$^{rd}$ image in Fig. 2(a)) is remarkable. The location of clusters getting fixed also occurs under equilibrium, but the number of clusters stopping changing is a unique behavior in ice growth scenario and does not happen under equilibrium. This behavior is better illustrated by the variation of cluster size distribution over time, see Fig. 2(b). In the first half of the simulation (blue to green lines), the cluster size curve goes upwards in a smooth way. After a certain time, small clusters start to shrink while large clusters keep growing, resulting in a "cliff" in the cluster size curve. As time goes further (orange to yellow lines), clusters on the "right" side of the cliff continue shrinking until disappear. On the contrary, clusters on the "left" side of the cliff (we will call them "*stable clusters*") continue growing, making the cliff steeper. This trend continues even after all small clusters have disappeared. Due to the continuous supplement of water molecules during ice growth, remaining stable clusters can grow simultaneously without any of them dissipating. As a result, the number of clusters does not change anymore.

Further we studied how the growth rate of ice affects the number density of stable clusters. In this paper, we use the inverse growth rate (IGR), the average time of adding a water molecule in unit area, as the measure of how fast ice grows (note that *higher* IGR means *slower* growth). We found that the inverse number density of stable clusters (= total area divided by the number of stable clusters) is positively linear dependent on IGR (see Fig. 2(c)). This relation can be qualitatively understood as faster growth increase the possibility of cluster formation and suppressing cluster shrinking. The linear dependence can be further explained by a simple model assuming free molecules in layer 1 are moving like random walks (see Supplementary Information Section 4).

It is worth noticing the significance of these stable clusters to real systems. Under realistic oversaturation, the growth rate of ice is much lower than those simulated in this work[36]. That means fewer stable clusters in a given area, or large "white space" between stable clusters. This resulting in the stable cluster distribution dominates the process of ice growth (see Supplementary Information Section 5), and determines its growth mechanism. This is the foundation of our discussion in the following sections.

## Growth mechanism of ice

Based on the picture of clusters, now we can discuss the growth mode of ice. Since cluster distribution is stable (yellow frame in Fig. 3), the growth mode of ice fully depends on the growth scheme of clusters. Theoretically, there are two possible schemes for cluster growth. In the first scheme (green frame), the clusters extend horizontally and keep the height of a single bilayer (1$^{st}$ sketch). The clusters continuously extend until merge together (2$^{nd}$ sketch) and form a full bilayer (3$^{rd}$ sketch), then the next bilayer of ice grows from here. In this paper, we name this the "bilayer-by-bilayer" mode of ice growth. In the second scheme (orange frame), the cluster extends both horizontally and vertically at the same time. Since ice is divided by bilayers in the vertical direction, the vertical growth of clusters means that new clusters (which have bilayer structure) form on top of existing ones (1$^{st}$-2$^{nd}$ sketch). Because clusters do not move after forming, new clusters will remain at their original positions after lower clusters merge into full bilayers (3$^{rd}$-4$^{th}$ sketch). In this paper we name this as the "cluster stacking" mode of ice growth. It is worth noticing that in this mode the clusters have similar behavior as droplets observed in experiments (exist during ice growth, have similar macroscopic appearance, look stable from surface, etc.).

It might be noticed that the two growth modes of ice here are somewhat similar to the ones in classical crystal growth theory[36,41]. It is yet unknown how to establish such theories for premelting surfaces though, due to complex structure and dynamics of QLL. By utilizing the concept of clusters and large scale molecular simulations, we can directly model the growth of ice without relying on empirical assumptions or external thermodynamic data.

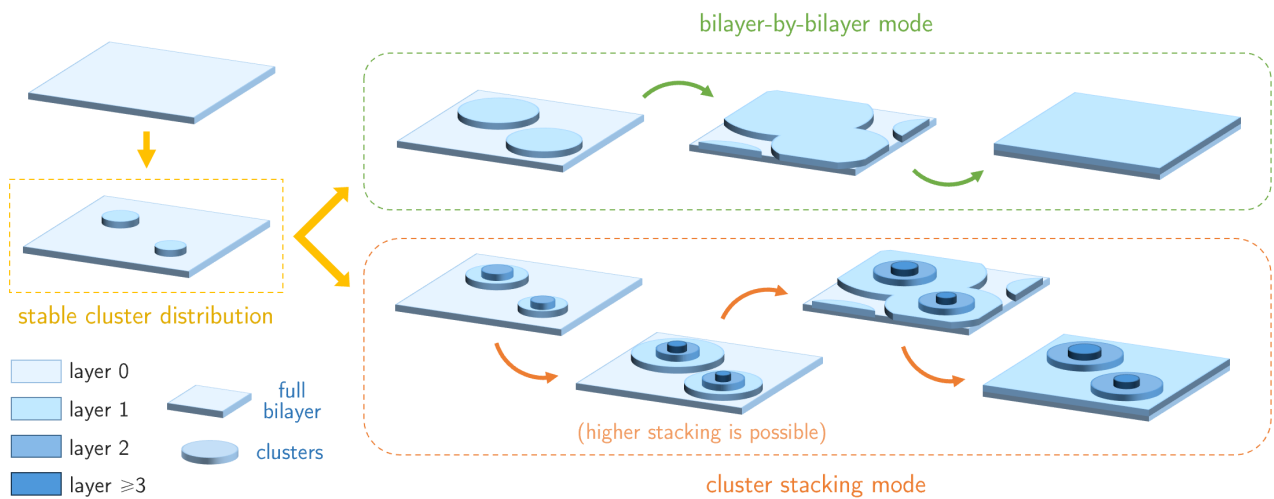

*Figure 3. Two possible schemes of cluster growth and ice growth. The illustration starts from the establishment of stable cluster distribution (in the yellow frame), and then two brunches split out. The one in green frame shows the bilayer-by-bilayer mode of ice growth, while the one in orange frame shows cluster stacking mode. Each pseudo-3D sketch represents a stage during the growth process. The large flat cuboids in sketches represent full bilayers, while disks represent clusters. Each bilayer or cluster stays in a single layer (layer 0/layer 1/...) and has the height of one bilayer. All bilayers and clusters are colored by the layer they belong to, see legends at the bottom-left corner (generally, deeper color means higher layer).*

To determine which growth mode actually happens in a given condition, we shall focus on the core difference between them: whether new clusters can form on top of existing ones. If new stable clusters can form on existing ones, the cluster stacking mode is preferred; otherwise the ice grows in the bilayer-by-bilayer mode. To this end, let's have a closer look at the growth process of clusters, see Fig. 4:

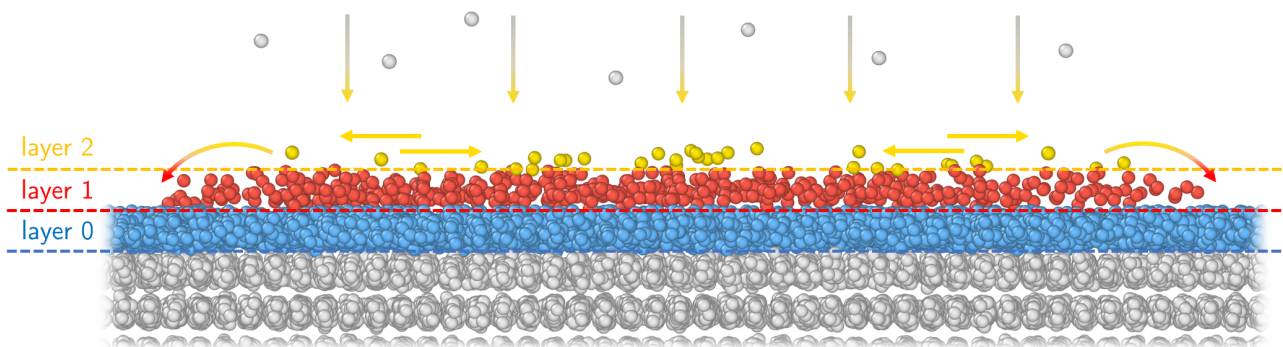

*Figure 4. Schematic of a cluster during ice growth. Dashed lines represent the boundaries of layers. Molecules in layer 0, 1, 2 are colored blue, red, and yellow, respectively. The cluster (mainly) locates in layer 1. White spheres on top of the figure represent new molecules coming to the surface. Arrows indicate the possible directions that molecules move, including coming to cluster, diffusing on the cluster, and transition from layer 2 to layer 1. Note that this figure is designed for illustration and doesn't faithfully represent real scenarios.*

Generally, new water molecules are continuously falling on the cluster during ice growth (vertical arrows in Fig. 4).

Roughly speaking, since clusters already have bilayer structure (see Supplementary Information Section 1), new molecules are unlikely to insert directly into the cluster (layer 1, colored red) but instead they accumulate above the cluster (layer 2, colored yellow). In the meantime, molecules may leave layer 2 by diffusing to the edge of the cluster (horizontal arrows) and jump into layer 1 (curved arrows). These two processes compete with each other and lead to a steady molecule number density in layer 2. In this paper the molecule number density in layer 2 is characterized by the layer concentration (LC) defined below:

$$LC(C) = \frac{N_{L2}(C)}{A(C)\,\sigma_{perfect}}$$

where $C$ represents a cluster, $A(C)$ is the area of the cluster, $N_{L2}(C)$ is the number of molecules in layer 2 on the cluster (yellow molecules in Fig. 4). $\sigma_{perfect}$ is a normalization factor equal to the molecule number density of perfect ice crystal (11.78 / nm$^2$, see Fig. 1(b)), so perfect crystal has 100% LC and other systems are scaled accordingly.

Intuitively higher LC should favor cluster formation in layer 2, thus favoring cluster stacking. Specifically, there should exist a *critical LC* (CLC), where new clusters will form in layer 2 after sufficient time if LC > CLC, and impossible to form otherwise. We found that the CLC is negatively related to the region size ($A(C)$ here), but tends to become a constant when the region is large enough to reach thermodynamic limit (see Supplementary Information Section 6). Furthermore, since in real scenarios stable clusters are far away from each other, stable clusters can grow very large so we can safely assume they have reached thermodynamic limit. That means we only need to consider CLC *under thermodynamic limit* when comparing it with LC.

Further we carried out the LC vs. CLC comparison on four different initial cluster sizes (~30, 40, 50, 60 nm, for the cluster in layer 1 of Fig. 3 and 4) under varied growth rates. This is done by tracing the evolution of LC during MD simulations, which are shown in Fig. 5. By tuning the initial structures (see Methods section) to ensure LC starts from estimated CLC (3.1%~3.3%, see Supplementary Information Section 6), the LC > CLC is simply indicated by an LC increase during simulations and vice versa. From Fig. 5 we can tell that generally low IGR (thus *high* growth rate) and large base cluster sizes favor cluster stacking. The former is easy to understand, while the latter is expected because time required for a new molecule diffusing to the edge of base cluster (see Fig. 4) depends on the cluster size.

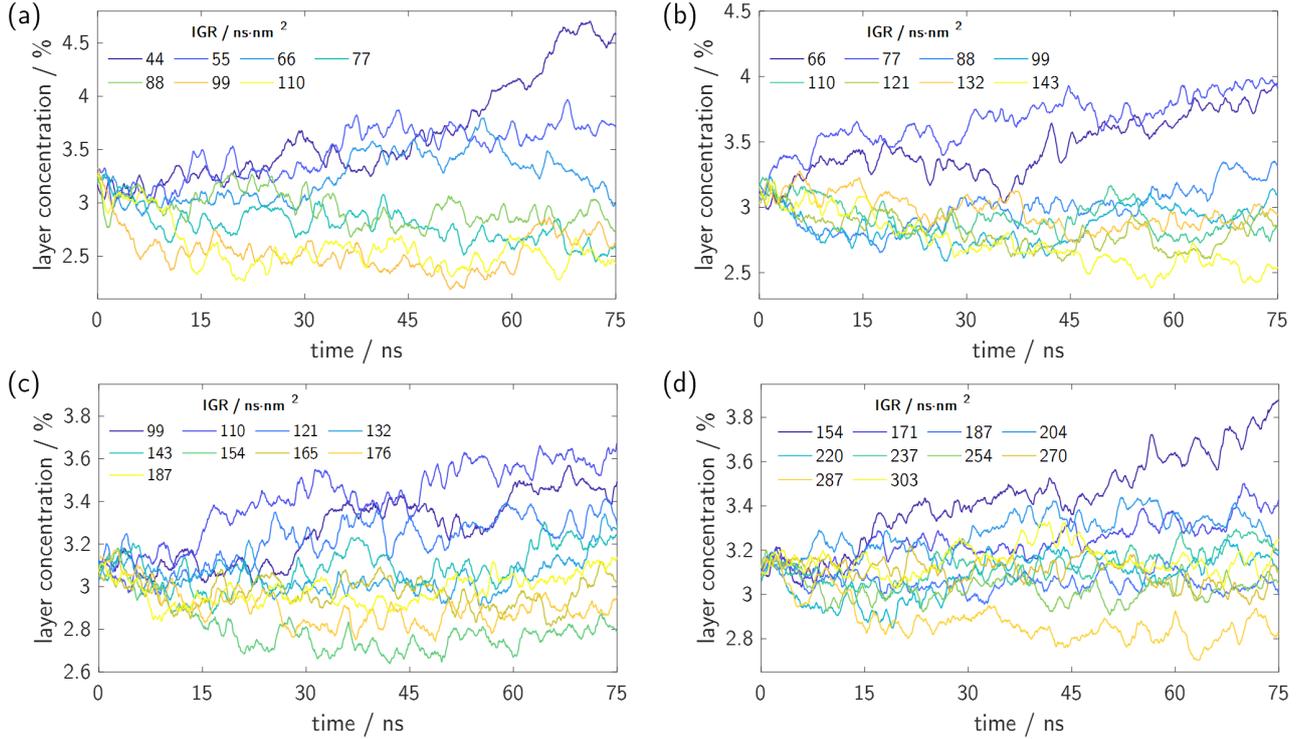

*Figure 5. Evolution of LC on base clusters during growth. Subfigures (a)~(d) demonstrate results in simulations with initial base cluster radius 30, 40, 50, and 60 nm, respectively. Each curve represents LC in a single simulation (with given IGR and base cluster size), as function of time. Curves are colored by their corresponding IGR (see legends in each subfigure). All curves are slightly smoothed for better visual appearance.*

One quantity we are interested in is $IGR_{max}$, the maximum IGR (thus *minimum* growth rate) for cluster stacking under a base cluster size. This is given by the boundary IGR between LC increase and decrease (remind generally lower IGR leads to higher LC). For example, from Fig. 5(a) we find $IGR_{max} \approx 66$ with base cluster size ~30 nm. However, the accuracy of determining $IGR_{max}$ decreases when the base cluster is large, due to the LC boundary becoming fuzzy (e.g. in Fig. 5(d)). This is caused by the longer time required for LC to reach steady, reflecting the challenge in studying QLL dynamics. Nevertheless, we can still get a rough estimate on $IGR_{max}$, which is not so quantitative but enough for our discussion below.

## Discussion

In the last section, we determined $IGR_{max}$ under given base cluster sizes. We may restate the result as we determined $r_{min}$, the *minimum base cluster size* required for cluster stacking under given *IGR* (since both low IGR and large base cluster size favor cluster stacking). Meanwhile, the *maximum* base cluster size in a real system is determined by the number density of stable clusters, which (as we know) is also determined by IGR. That points out the way to determine the feasibility of cluster stacking, namely comparing "minimum and maximum" under the same IGR:

(1) $r_{min}$, minimum base cluster size required for cluster stacking;

(2) $r_{max}$, maximum base cluster size, derived from the number density of stable clusters.

To form cluster stacks, the base cluster size must be larger than $r_{min}$; meanwhile, the base cluster size on ice surface is limited by size $r_{max}$. Therefore, $r_{min} > r_{max}$ means cluster stacking is impossible and the ice grows in bilayer-by-bilayer mode; and $r_{min} < r_{max}$ means cluster stacking can occur and the ice grows in cluster stacking mode. With the $r_{max}/r_{min}$ ratio going higher, more layers can exist in cluster stacks. Fig. 6 shows both sizes under IGR range investigated in this paper. It's clear that $r_{min}$ is larger here, indicating a bilayer-by-bilayer growth mode in this IGR range (note that the "crossover" in ultra-low IGR region is not physically meaningful; cluster stacking would not occur there, because base clusters in that region are too small to form cluster stacks).

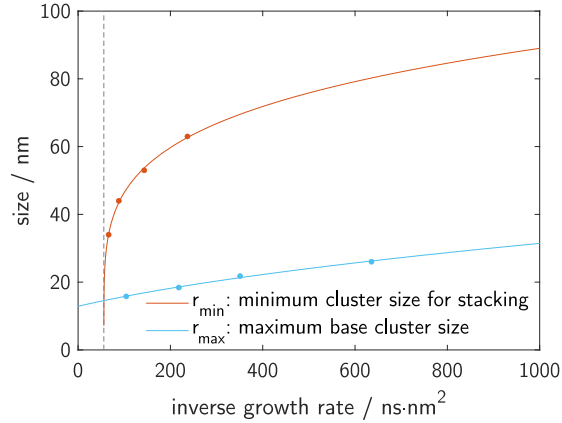

Figure 6. Comparison of two sizes. Dots represent data points we obtained from simulations (Fig. 2(c) and Fig. 5, respectively), and curves are fittings of them ($r_{min}$: $(ax+b)^c$; $r_{max}$: $(ax+b)^{0.5}$). The $r_{max}$, maximum base cluster size, is equal to the <u>half</u> of average distance between stable clusters (since we characterize cluster size by radius). The $r_{min}$, minimum cluster size for stacking, is equal to the <u>final</u> base cluster size of corresponding simulations (34, 44, 53, 63 nm, respectively), so it's slightly larger than the initial sizes used in Section 3.3 (30, 40, 50, 60 nm). Such difference is minor and doesn't affect our discussion.

Unfortunately we cannot directly determine the growth mode in realistic scenarios, because the IGR is way larger than those investigated here and unreachable for today's MD simulations. However, we may touch this by how molecules in layer 2 diffuse on clusters, since it's directly related to $r_{min}$. For full bilayers, molecules above the bilayer diffuse like random walks (see Supplementary Information Section 4), resulting in a $(ax+b)^{0.5}$ behavior for $r_{max}$ (blue curve in Fig. 6). We may fit $r_{min}$ with a similar model $(ax+b)^c$, resulting in c = 0.21 (orange curve in Fig. 6). The smaller exponent indicates that the diffusion of molecules on clusters is more convoluted than random walk, which is probably caused by the barrier from edges. Furthermore, we speculate such effect might persist to the thermodynamic limit, because the barrier always exists and the "diffusion" in a liquid-like system is global. Therefore, the small exponent may persist to large IGR scenarios, resulting in $r_{min} < r_{max}$ and cluster stacking happening in real systems. Confirming or denying such speculation requires a systematic investigation of how water molecules transport in clusters, which will significantly enhance our understanding of QLL.

Finally, it is worth mentioning that some other factors may also affect the growth mode of ice. An obvious one is temperature, which is suggested to cause structural transition in QLL[21,29]. Such transition may lead to an abrupt change in the diffusion behavior of molecules, as well as stable cluster distribution. Another possible factor is defects, which may induce or restrain the formation of clusters, and impact the diffusion. Some experiments found that the distribution of defects and droplets are correlated on ice surface[42], strengthening such speculations.

## Conclusion

In this paper, we report a systematic molecular-level study of QLL during ice growth, under large spatial scales (70~280 nm) using mW water model. We discovered the spontaneous formation of clusters, and simulated their distribution and evolution. Specifically, a stable cluster distribution exists during ice growth, with the number density of stable clusters linearly dependent on IGR. Based on this picture we further purpose two possible growth modes of ice, namely bilayer-by-bilayer and cluster stacking. The actual growth mode is determined by calculating LC evolution on clusters and comparing "$r_{min}$ and $r_{max}$", and we found the bilayer-by-bilayer mode being real under simulated high growth rates. However, the growth mode under realistic low growth rates is yet unrevealed and we suggest potential researches in the future. Overall, the concept of clusters and their distribution points out a way of connecting molecular-level dynamics and macroscopic growth of ice, making a step forward to the full understanding of QLL in ice growth.

## Methods

### Model construction

**Unit cell.** Model construction begins from building the unit cell. The lattice constant of $I_h$ ice used in this work is derived from simulating bulk ice crystal with a large supercell under NPT ensemble, resulting in a = b = 0.4427 nm, c = 0.72028 nm. The unit cell of $I_h$ ice is constructed with this lattice constant, then the locations of molecules are optimized by minimizing the energy. Then this optimized unit cell is duplicated to construct various models in this work.

**QLL under equilibrium.** The simulation begins with a perfect $I_h$ ice crystal, having (0001) surface exposed on the top in z direction. The model size in x and y direction is ~140 x 140 nm. The model has 6 bilayers in z direction in total, with the bottom 2 bilayers fixed during simulation and top 4 bilayers movable (since QLL generally only have 1~2 bilayers thick, 4 movable bilayers would be enough). ~3 nm of vacuum is preserved above the surface (this is more than enough since the mW force field is short-ranged with cutoff of 0.432 nm[40]), and a virtual reflective wall is set above the vacuum layer to avoid losing molecules. Periodic boundary conditions are imposed on x and y directions.

**QLL during ice growth.** First we constructed a perfect $I_h$ ice crystal with sufficient bilayers in z direction and (0001) surface exposed on both sides. The model is then relaxed by NVT simulation of 1 ns. After that, we extracted 4.5 bilayers (0.5 means a half-bilayer) from the top of the model as the initial structure of our simulation. *We shall point out that*, we believe the procedure above is not necessary, and anyone who wants to reproduce may simply use perfect ice crystals as their initial structures (like we did in the simulation under equilibrium). However, such a procedure (though overcomplicated) is unlikely to affect our conclusion, since the number of stable clusters (an integer) is relatively robust and our conclusion doesn't rely on very accurate numbers either.

Anyway, after the construction above, we further removed some molecules from the surface to make sure no clusters may form at the beginning of simulation. Specifically, the initial structure should have LC << CLC in layer 1, and remaining molecules in layer 1 should be as scattered as possible. In simulations the bottom 1.5 bilayers are fixed and the top 3 layers are movable. ~6 nm of vacuum is preserved above the surface, but without a reflection layer.

**Growth mechanism of ice.** Model construction also begins from perfect $I_h$ ice crystals same as the NVT simulation, but larger in x and y directions (~210 x 210 nm). Then the crystal is relaxed by NVT simulation of 2.5 ns. After that, most molecules in layer 0 and layer 1 are deleted, only remaining a circle filled in the middle of the model. The circle may have a radius of 30~60 nm (see main text) and would be the base cluster. Further some molecules in layer 1 of the circle are deleted to adjust LC on the cluster, and the model is relaxed again for 10~25 ns. This process may be repeated multiple times to make sure the initial structure is sufficiently relaxed. After the relaxation some clusters may appear in layer 0 outside the base cluster, and we deleted them to avoid them touching the base cluster during simulations. ~6 nm of vacuum is preserved above the surface, and a reflection layer exists above the vacuum layer.

## Simulation settings

**General settings.** MD simulations are performed with LAMMPS[43] software, accelerated by KOKKOS[44] or INTEL[45] package. All simulations are done under temperature 273.9 K, using Nose-Hoover thermostat[46,47] with relaxation time of 1.25 ps. The mW water model[40] is used in this work. The timestep of 5 fs is used in all simulations[20]. Most analysis and visualization are performed with Matlab[48] and Ovito[49] software.

**QLL under equilibrium.** The simulation is carried out under NVT ensemble for 250 ns. A snapshot of intermediate structure is made every 0.25 ns.

**QLL during ice growth.** To simulate ice growth, molecules are added to the system at 3.5~6 nm above the surface every given time interval (this interval determines IGR). Newly added molecules are given initial velocities of 0.1~0.2 nm/ps pointing to the surface, so most (but not all) molecules will fly to the surface. Note that the IGR reported in the main text is calculated by the *net* change of molecule numbers, i.e. molecules leaving the model are subtracted. For reference, four data points in Fig. 3(c) corresponds to molecule adding interval of 5.25, 10.5, 16, 26.25 ps, respectively, in a 136.151 x 139.35 nm model. Each simulation is performed until the stable cluster distribution is established.

**Growth mechanism of ice.** Similar to Section 2.2, but molecules are added at 3~4.5 nm above the surface (such difference is arbitrary, and we believe it's not important because mW force-field is short-ranged), and a snapshot is made every 0.05 ns.

## Data analysis

**Dividing layers.** The lower boundary of layer 0 is the bisector of first two bilayers in the initial structure. The layer 0 is one bilayer thick (0.36 nm), and layer 1 generally represents the whole region above layer 0. In the context of requiring more than two layers (Fig. 3, Fig. 4), each layer is one bilayer thick, and the region of each layer is determined successively from layer 0. The region corresponding to each layer is fixed in the whole simulation.

**Cluster analysis.** Cluster analysis is performed with distance-based neighbor criterion: two molecules with distance < 0.45 nm belong to the same cluster. This is done with the "cluster analysis" modifier in Ovito. Only molecules in layer 1 are considered.

**Cluster tracing (Fig. 1(d)).** To get Fig. 1(d) we need to know where a cluster in a certain snapshot comes from (e.g. "cluster B in snapshot 498 comes from cluster A in snapshot 497"). This is done in the following procedure:

(1) Calculate the center of each cluster in snapshot [n] (we will call them $A_1, A_2,...$) and snapshot [n+1] (we will call them $B_1, B_2,...$). Only clusters with >50 molecules are considered.

(2) Calculate pairwise distance matrix between $A_1, A_2,...$ and $B_1, B_2,...$.

(3) Select the smallest element in the distance matrix. If it's larger than 10 nm, go to (6); else go to (4).

(4) Find the corresponding $A_x$ and $B_y$ for this element, and we will claim "cluster $B_y$ in snapshot [n+1] comes from cluster $A_x$ in snapshot [n]".

(5) Set this element to +infinity, and go back to (3).

(6) If a cluster in snapshot [n+1] doesn't come from any cluster in [n], we will claim it's newly evolved at this snapshot.

(7) If a cluster in snapshot [n] doesn't have any corresponding cluster in [n+1], we will claim it disappears in [n+1].

The choice of cutoff ("10 nm" in (3) and "50 molecules" in (1)) is theoretically arbitrary but should be reasonable to get meaningful results.

**Distribution of water and ice (Supplementary Information Section 3).** The method described in this article[50] (called "identify diamond structure" in Ovito) is used to determine if a molecule belongs to water or ice (hexagonal / cubic) structure. This method is originally designed for Si and Ge systems, but we found it also works well for mW water, including identifying structures in QLL. Note that the 1st and 2nd nearest neighbors of ice molecules are also categorized as ice molecules in this work (themselves are "on the lattice site" but their neighbors may not).

**Stable cluster distribution.** The stable cluster distribution is considered established when all of the following criteria are met: (1) every stable cluster has >400 molecules; (2) every other cluster has <100 molecules; (3) cluster distribution doesn't change after this time (except merging due to cluster growth). Due to thermal fluctuations, the time when cluster distribution reaches stable (used in Fig. S4) has an uncertainty of ~10 ns.

**LC on clusters.** LC is calculated in the layer above the base cluster (layer 2 in Fig. 4). The area occupied by the base cluster is calculated by making a boundary polygon for molecules in the base cluster. To calculate the boundary polygon, we project the molecules to XY plane and construct alpha shapes from them. All distinct connected alpha shapes are sorted by their alpha value, and the shape with alpha value at the lower quartile is used to extract the boundary polygon. The whole process is done with the function "boundary" in Matlab.

## Acknowledgments

We thank Prof. Ying Jiang for introducing us to this research field (premelting of ice), Haotian Zheng and Xinmeng Liu for help in preliminary investigations, along with Tonghuan Jiang, Yilin Chen and Prof. Ji Chen for helpful discussions. We also thank Prof. Ji Chen for providing computational resources. This work was supported by the National Natural Science Foundation of China under Grant No. 11974024.

## References

1. FARADAY, M. R. On certain conditions of freezing water. *Athenaeum*. **1181**, 640-641 (1850).
2. Dash, J. G., Fu, H. & Wettlaufer, J. S. The premelting of ice and its environmental consequences. *Rep. Prog. Phys.* **58**, 115-167 (1995).
3. Bartels-Rausch, T. et al. A review of air-ice chemical and physical interactions (AICI): liquids, quasi-liquids, and solids in snow. *Atmos. Chem. Phys.* **14**, 1587-1633 (2014).
4. Thomas, J. L. et al. Modeling chemistry in and above snow at Summit, Greenland - Part 2: Impact of snowpack chemistry on the oxidation capacity of the boundary layer. *Atmos. Chem. Phys.* **12**, 6537-6554 (2012).
5. Rempel, A. W. Formation of ice lenses and frost heave. *J. Geophys. Res.: Earth Surf.* **112**, F02S21 (2007).
6. Yu, F., Guo, P., Lai, Y. & Stolle, D. Frost heave and thaw consolidation modelling. Part 1: A water flux function for frost heaving. *Can. Geotech. J.* **57**, 1581-1594 (2020).
7. Kim, K. & Park, M. J. Ice-assisted synthesis of functional nanomaterials: The use of quasi-liquid layers as nanoreactors and reaction accelerators. *Nanoscale*. **12**, 14320-14338 (2020).
8. Chen, D., Gelenter, M. D., Hong, M., Cohen, R. E. & McKinley, G. H. Icephobic surfaces induced by interfacial nonfrozen water. *ACS Appl. Mater. Interfaces*. **9**, 4202-4214 (2017).
9. Sánchez, M. A. et al. Experimental and theoretical evidence for bilayer-by-bilayer surface melting of crystalline ice. *Proc. Natl. Acad. Sci. U. S. A.* **114**, 227-232 (2017).
10. Bluhm, H., Ogletree, D. F., Fadley, C. S., Hussain, Z. & Salmeron, M. The premelting of ice studied with photoelectron spectroscopy. *J. Phys.: Condens. Matter*. **14**, L227 (2002).
11. Dosch, H., Lied, A. & Bilgram, J. H. Glancing-angle X-ray scattering studies of the premelting of ice surfaces. *Surf. Sci.* **327**, 145-164 (1995).
12. Mizuno, Y. & Hanafusa, N. Studies of surface properties of ice using nuclear magnetic resonance. *Le Journal de Physique Colloques*. **48**, 511-517 (1987).
13. Golecki, I. & Jaccard, C. Intrinsic surface disorder in ice near the melting point. *J. Phys. C: Solid State Phys.* **11**, 4229 (1978).
14. Lecadre, F., Kasuya, M., Kanno, Y. & Kurihara, K. Ice Premelting Layer Studied by Resonance Shear Measurement (RSM). *Langmuir*. **35**, 15729-15733 (2019).
15. Miyato, Y., Otani, K., Maeda, M., Nagashima, K. & Abe, M. Investigating ice surfaces formed near the freezing

point in the vapor phase via atomic force microscopy. *Jpn. J. Appl. Phys.* **58**, (2019).

16. Mitsui, T. & Aoki, K. Fluctuation spectroscopy of surface melting of ice with and without impurities. *Phys. Rev. E.* **99**, 10801 (2019).

17. Smit, W. J. & Bakker, H. J. The Surface of Ice Is Like Supercooled Liquid Water. *Angew. Chem., Int. Ed.* **56**, 15540-15544 (2017).

18. Li, H. et al. Water Mobility in the Interfacial Liquid Layer of Ice/Clay Nanocomposites. *Angew. Chem., Int. Ed.* **60**, 7697-7702 (2021).

19. Fletcher, N. H. Surface structure of water and ice. *Philos. Mag.* **18**, 1287-1300 (1968).

20. Pickering, I., Paleico, M., Sirkin, Y. A. P., Scherlis, D. A. & Factorovich, M. H. Grand Canonical Investigation of the Quasi Liquid Layer of Ice: Is It Liquid? *J. Phys. Chem. B.* **122**, 4880-4890 (2018).

21. Benet, J., Llombart, P., Sanz, E. & MacDowell, L. G. Premelting-Induced Smoothening of the Ice-Vapor Interface. *Phys. Rev. Lett.* **117**, (2016).

22. Limmer, D. T. & Chandler, D. Premelting, fluctuations, and coarse-graining of water-ice interfaces. *J. Chem. Phys.* **141**, 18C505 (2014).

23. Watkins, M. et al. Large variation of vacancy formation energies in the surface of crystalline ice. *Nat. Mater.* **10**, 794-798 (2011).

24. Bishop, C. L. et al. On thin ice: surface order and disorder during pre-melting. *Faraday Discuss.* **141**, 277-292 (2009).

25. Conde, M. M., Vega, C. & Patrykiejew, A. The thickness of a liquid layer on the free surface of ice as obtained from computer simulation. *J. Chem. Phys.* **129**, 14702 (2008).

26. Pan, D. et al. Surface energy and surface proton order of ice Ih. *Phys. Rev. Lett.* **101**, 155703 (2008).

27. Sibley, D. N., Llombart, P., Noya, E. G., Archer, A. J. & MacDowell, L. G. How ice grows from premelting films and water droplets. *Nat. Commun.* **12**, 239 (2021).

28. Pan, D., Liu, L., Slater, B., Michaelides, A. & Wang, E. Melting the Ice: On the Relation between Melting Temperature and Size for Nanoscale Ice Crystals. *ACS Nano.* **5**, 4562-4569 (2011).

29. Noya, E. G., Sibley, D. N., Archer, A. J., MacDowell, L. G. & Llombart, P. Rounded Layering Transitions on the Surface of Ice. *Phys. Rev. Lett.* **124**, 65702 (2020).

30. Kroes, G. Surface melting of the (0001) face of TIP4P ice. *Surf. Sci.* **275**, 365-382 (1992).

31. Qiu, Y. & Molinero, V. Why Is It so Difficult to Identify the Onset of Ice Premelting? *J. Phys. Chem. Lett.* **9**, 5179-5182 (2018).

32. Beaglehole, D. & Nason, D. Transition layer on the surface on ice. *Surf. Sci.* **96**, 357-363 (1980).

33. Buch, V., Groenzin, H., Li, I., Shultz, M. J. & Tosatti, E. Proton order in the ice crystal surface. *Proc. Natl. Acad. Sci. U. S. A.* **105**, 5969-5974 (2008).

34. Slater, B. & Michaelides, A. Surface premelting of water ice. *Nat. Rev. Chem.* **3**, 172-188 (2019).

35. Libbrecht, K. G. Physical Dynamics of Ice Crystal Growth. *Annu. Rev. Mater. Res.* **47**, 271-295 (2017).

36. Libbrecht, K. G. The physics of snow crystals. *Rep. Prog. Phys.* **68**, 855-895 (2005).

37. Nagata, Y. et al. The Surface of Ice under Equilibrium and Nonequilibrium Conditions. *Acc. Chem. Res.* **52**, 1006-1015 (2019).

38. Asakawa, H., Sazaki, G., Nagashima, K., Nakatsubo, S. & Furukawa, Y. Two types of quasi-liquid layers on ice crystals are formed kinetically. *Proc. Natl. Acad. Sci. U. S. A.* **113**, 1749-1753 (2016).

39. Sazaki, G., Zepeda, S., Nakatsubo, S., Yokomine, M. & Furukawa, Y. Quasi-liquid layers on ice crystal surfaces are made up of two different phases. *Proc. Natl. Acad. Sci. U. S. A.* **109**, 1052-1055 (2012).

40. Molinero, V. & Moore, E. B. Water Modeled As an Intermediate Element between Carbon and Silicon. *J. Phys. Chem. B.* **113**, 4008-4016 (2009).

41. Saito, Y. *Statistical Physics of Crystal Growth*. (WORLD SCIENTIFIC, 1996).

42. Sazaki, G., Asakawa, H., Nagashima, K., Nakatsubo, S. & Furukawa, Y. How do Quasi-Liquid Layers Emerge from Ice Crystal Surfaces? *Cryst. Growth Des.* **13**, 1761-1766 (2013).

43. Thompson, A. P. et al. LAMMPS - a flexible simulation tool for particle-based materials modeling at the atomic, meso, and continuum scales. *Comput. Phys. Commun.* **271**, 108171 (2022).

44. Edwards, H. C., Trott, C. R. & Sunderland, D. Kokkos: Enabling manycore performance portability through polymorphic memory access patterns. *J. Parallel Distr. Com.* **74**, 3202-3216 (2014).

45. Michael Brown, W., Carrillo, J. Y., Gavhane, N., Thakkar, F. M. & Plimpton, S. J. Optimizing legacy molecular dynamics software with directive-based offload. *Comput. Phys. Commun.* **195**, 95-101 (2015).

46. Nosé, S. A unified formulation of the constant temperature molecular dynamics methods. *J. Chem. Phys.* **81**, 511-519 (1984).

47. Hoover, W. G. Canonical dynamics: Equilibrium phase-space distributions. *Phys. Rev. A.* **31**, 1695-1697 (1985).

48. The MathWorks Inc. MATLAB. version 9.10.0 (R2021a). Natick, Massachusetts; 2021.

49. Stukowski, A. Visualization and analysis of atomistic simulation data with OVITO-the Open Visualization Tool. *Model. Simul. Mater. Sc.* **18**, 15012 (2009).

50. Maras, E., Trushin, O., Stukowski, A., Ala-Nissila, T. & Jónsson, H. Global transition path search for dislocation formation in Ge on Si(001). *Comput. Phys. Commun.* **205**, 13-21 (2016).

# Supplementary Information

For article: *Premelting layer during ice growth: role of clusters*

Shifan Cui[1], Haoxiang Chen[2], Zhengpu Zhao[1]

## Contents

1 International Center for Quantum Materials, School of Physics, Peking University, Beijing, China.
2 School of Physics, Peking University, Beijing, China.

# 1 Vertical distribution of molecules

In the "QLL under equilibrium" section of main text we studied the horizontal distribution of molecules in layer 1, and showed the existence of clusters. To further clarify the structure of these clusters, we calculated the vertical distribution of molecules (also in this simulation), at two typical time points: (1) 1 ns, where no clusters have formed; (2) 250 ns, where several large clusters exist. The results are shown in Fig. S1.

Fig. S1(a) shows the vertical molecule distribution profile at two time points, along with the profile of initial structure (perfect crystal) as a comparison. It can be seen that the molecule distribution has almost reached equilibrium after 1 ns, and the shape of the profile agrees with previous studies[1,2]. However, if we look closely into the region of layer 1 (the inset of Fig. 1(a)), we can notice a significant difference between the profile of two time points (1 ns vs. 250 ns). Namely, the molecule count decays monotonically toward the surface at t = 1 ns (orange line), but a peak at ~1.2 nm exists at t = 250 ns (yellow line). Further analysis shows that this is related to the existence of clusters. Fig. S1(b) explicitly illustrated the vertical profile of layer 1, and decomposed the contribution from clusters and "free" molecules (molecules don't belong to any cluster) for t = 250 ns (all molecules are free at t = 1 ns). The profile of free molecules is always monotonically decreasing, but a peak exists for the profile of clusters (the "t = 250 ns" graph of Fig. S1(b), orange line). Furthermore, we noticed that the profile of clusters and the equilibrated first bilayer (layer 0) are very similar if we plot them together (bottom graph). This similarity indicates that the clusters actually have the structure of bilayers. Meanwhile, this also means that the clusters are actually "located" in layer 1, and relatively isolated from layer 0 (this can be confirmed by the vertical profile of clusters, which is ~0 at the boundary of two layers). Therefore, both the concept of clusters, and the division between two layers, are well-behaved.

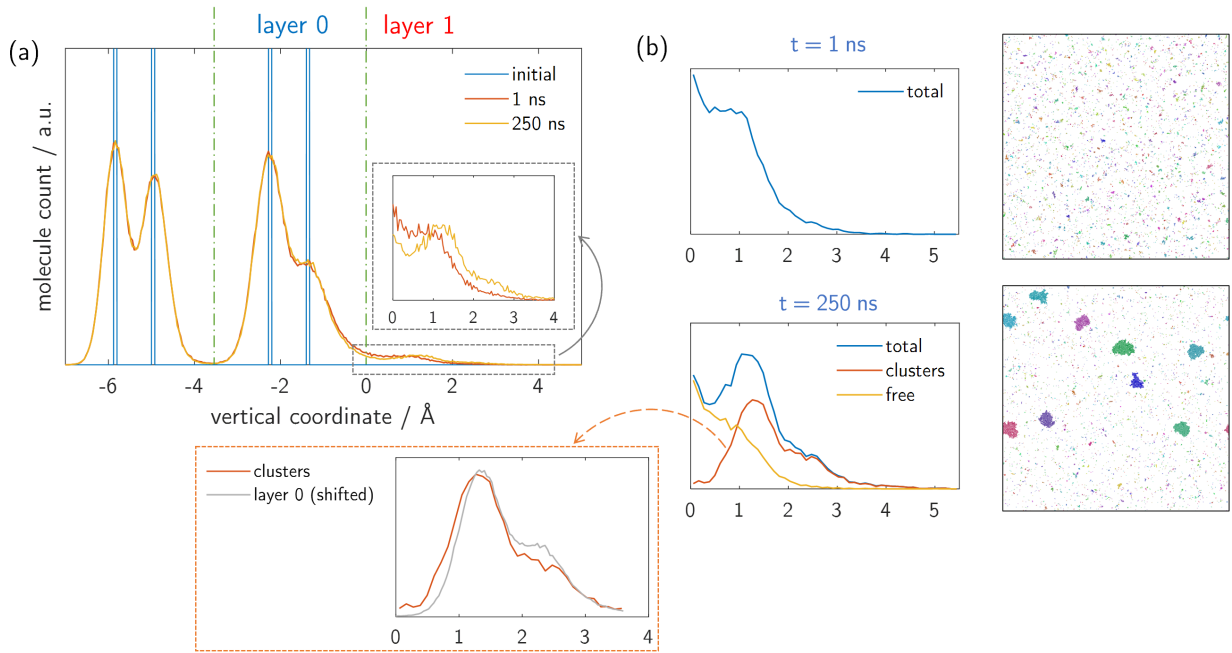

Figure S1. The vertical molecule distribution profile. (a) The vertical profile of initial structure (blue), structure at 1 ns (orange) and 250 ns (yellow), in vertical coordinate range -7~5 Å. The zero point of vertical coordinate is the boundary between layer 0 and layer 1. The inset is a magnification of curves in the 0~4 Å region. The green vertical lines are boundaries of layers. (b) The vertical profile of structure at 1 ns (upper) and 250 ns (lower), in vertical coordinate range 0~5 Å. The meanings of both axes are same as (a). The horizontal distributions of molecules in layer 1 at two times are shown aside for reference. For decomposing the profile of clusters and free molecules, molecule "groups" with >50 molecules are treated as clusters. The graph at bottom shows the comparison between the profile of clusters (from the 250 ns graph in (b)) and layer 0 (from the 250 ns curve in (a), shifted to draw both curves together).

## 2 Clusters are unstable in small model sizes

In the "QLL under equilibrium" section of main text we show that small clusters are unstable. In small simulation models, there are not enough molecules in layer 1, prohibiting large clusters to form. The remaining small clusters tend to dissipate quickly, thus are unable to be observed consistently. This is probably the reason why clusters are not reported before.

This argument can be confirmed directly by an MD simulation with a much smaller model (~20 x 20 nm horizontal, ~28000 molecules). All other simulation settings are kept same as the "QLL under equilibrium" section of main text. Results are shown in Fig. S2.

Fig. S2(a) shows the horizontal distribution of molecules in layer 1 at three different times. It is possible to form clusters in such small models, as shown in the middle image (t = 150 ns). However, the cluster breaks up later (see the right image, t = 225 ns), meaning that such clusters are unstable and cannot be consistently observed. Fig. S2(b) shows the change of largest cluster size during simulation. Some indication of forming clusters can be seen (peaks in the curve), but they are too small (<200 molecules) thus cannot exist for a long time.

It is worth mentioning that, it is still possible to observe stable clusters in small models. As demonstrated in the main text later (and Supplementary Information Section 6), layer concentration (LC) affects cluster formation, and increasing LC will help the formation of clusters in such small models. This can be achieved by adding extra molecules on the surface of initial structure, so more molecules will stay in layer 1 at equilibrium. However, the number of molecules added must be carefully controlled, otherwise clusters may contact each other (or with the periodic mirror of itself).

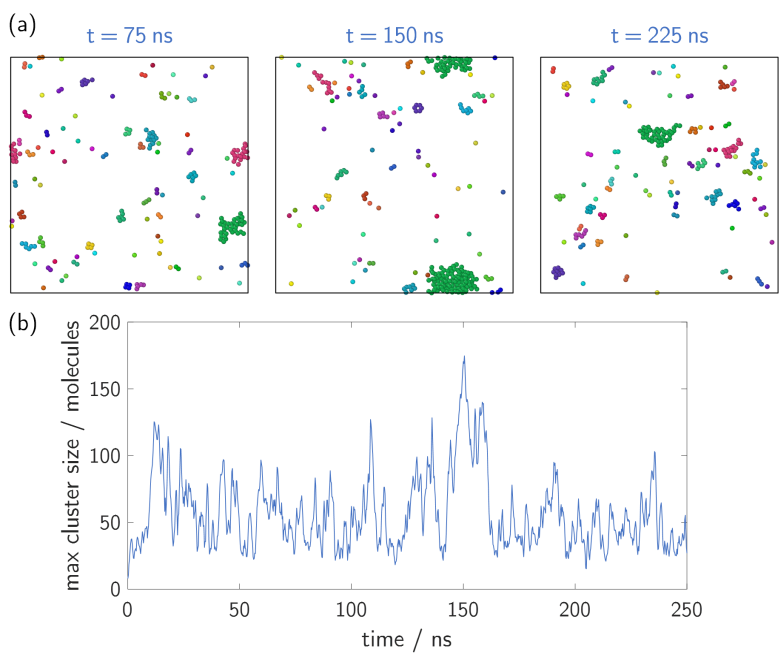

*Figure S2. Evolution of clusters in a 20 x 20 nm model. (a) horizontal distribution of molecules in layer 1, at three different times of simulation (75 ns, 150 ns, 225 ns). Molecules are colored by the clusters they belong to. Color choices of different images are not related. (b) maximum cluster size in layer 1 during simulation, as function of time. The curve is slightly smoothed for a better visual appearance.*

# 3 Distribution of water and ice in QLL

Fig. S3 demonstrates the distribution of water (more accurately, molecules with liquid-like structure) and ice (molecules with solid-like structure) in QLL. Fig. S3(a) shows the water-ice distribution of several clusters. They are composed of both water and ice, with stacking faults occurring occasionally. Fig. S3(b) shows the same thing but for the whole layer 0, which shows similar behavior as that of clusters. Such a mixed distribution agrees with results in previous study[3]. However, this is not true for the region in layer 0 covered by clusters. The shadowed area in Fig. 3(b) shows the location of these clusters, and most of these area is composed of ice (colored yellow/blue) in layer 0. That is to say, clusters are mostly "sat" on solid ice rather than liquid water.

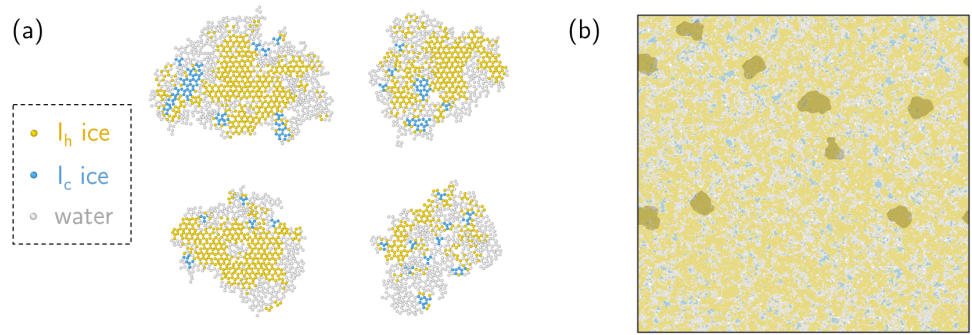

*Figure S3. Distribution of water and ice in QLL (clusters & layer 0). Hexagonal ice, cubic ice, and water molecules are colored yellow, blue, and grey, respectively (see Methods section of main text for how they're categorized). The structure is taken from the end of simulation in the "QLL under equilibrium" section of main text (t = 250 ns). (a) four typical clusters. (b) layer 0 as a whole; shadows indicate region covered by clusters (note that clusters themselves are not shown in (b), since they're not in layer 0).*

# 4 A simple model on the number density of stable clusters

The linear dependence between number density of stable clusters and inverse growth rate (IGR) can be explained by a simple model. First note that molecules in layer 1 can be divided into two types: (1) molecules in clusters, and (2) free molecules (molecules not in any cluster). We will make the following assumptions about these two types of molecules:

- Free molecules move as random walking in layer 1 until they touch any cluster;
- After touching a certain cluster, free molecules will become molecules in that cluster;
- Molecules in clusters always stay in the same cluster.

We will further make the following simplification:

- Area occupied by clusters in layer 1 is small compared to the system size (this is usually true, see Supplementary Information Section 5), so (almost) all molecules newly come to the surface are free molecules;
- The growth of clusters is slow (compared to the interval of molecules coming to surface), so the area of "free region" (region not covered by clusters) can be considered as a constant.

Since the amount of clusters does not change, the LC in free region should roughly be a constant (if LC increases to too high, new clusters will form in free region; if LC decreases to too low, existing clusters will dissipate). Since the area of free region is considered constant, the number of free molecules has to be constant. That is to say, the following two rates are equal:

**(1)** Rate of new free molecules added;

**(2)** Rate of free molecules touching clusters.

(1) is (by simplification) equal to the rate of new molecules coming to the surface, or growth rate. So the inverse of (1) is IGR. The inverse of (2) is the average time required for a free molecule to touch a cluster. Since free molecules are assumed to do random walks, this average time is proportional to $d^2$, where $d$ is the average distance between a random point in free region and the nearest cluster. Since clusters are (roughly) uniformly located, $d$ is further proportional to the average distance between clusters. Then $1/d^2$ is proportional to the number of clusters in unit area, so $d^2$ itself is proportional to inverse number density of clusters. In other words, the inverse of (2) is proportional to the inverse number density of clusters. Since (1) is equal to (2), the inverse number density of clusters is proportional to IGR. The intercept in Fig. 3(c) of main text represents the finite size of clusters, so the number density of clusters cannot be infinite. However, such effect is negligible in realistic scenarios (IGR is very large).

It's necessary to point out that the real world is more complicated. The "mean free path" of free molecules in layer 1 is actually very short, due to frequent exchanges of molecules between layer 0 and layer 1. So the free molecules are not really doing random walks. More likely, the "random walk" of free molecules in layer 1 represents the diffusion of molecules in layer 0, which arguably shows $t \sim d^2$ behavior if traditional diffusion theory is applicable.

# 5 The stable distribution of clusters lasts long during ice growth

In the "QLL during ice growth" section of main text, we have known about the stable cluster distribution during ice growth. In this section, we will discuss how long does the "stable distribution" hold in the whole process of growth. Here we assume that the ice grows in bilayer-by-bilayer mode (see Fig. 3 of main text; for cluster stacking mode the stable distribution arguably holds at all times). In this mode, the stable distribution establishes when the number of clusters stops changing, and lasts until clusters start to contact each other. Fig. S4(a) shows the horizontal map of molecules in layer 1, right after the establishment of stable distribution. It can be seen that, with IGR increases, the number of molecules in layer 1 at this moment decreases. Fig. S4(b) demonstrates this quantitively by showing LC corresponding to (a). Therefore, under high IGR in realistic scenarios, the stable distribution starts holding when LC is very low (<10%, at most). On the other hand, since clusters are (roughly) uniformly located, the LC when clusters start contacting is mostly determined by how they arrange in the 2D plane. In a typical example (simple square arrangement), the LC at this time is $\pi/4 \approx 78.5\%$. Therefore, during the period of growing a full bilayer, the stable distribution holds for ~70% of total time. So we can say the stable cluster distribution dominates the whole process of ice growth.

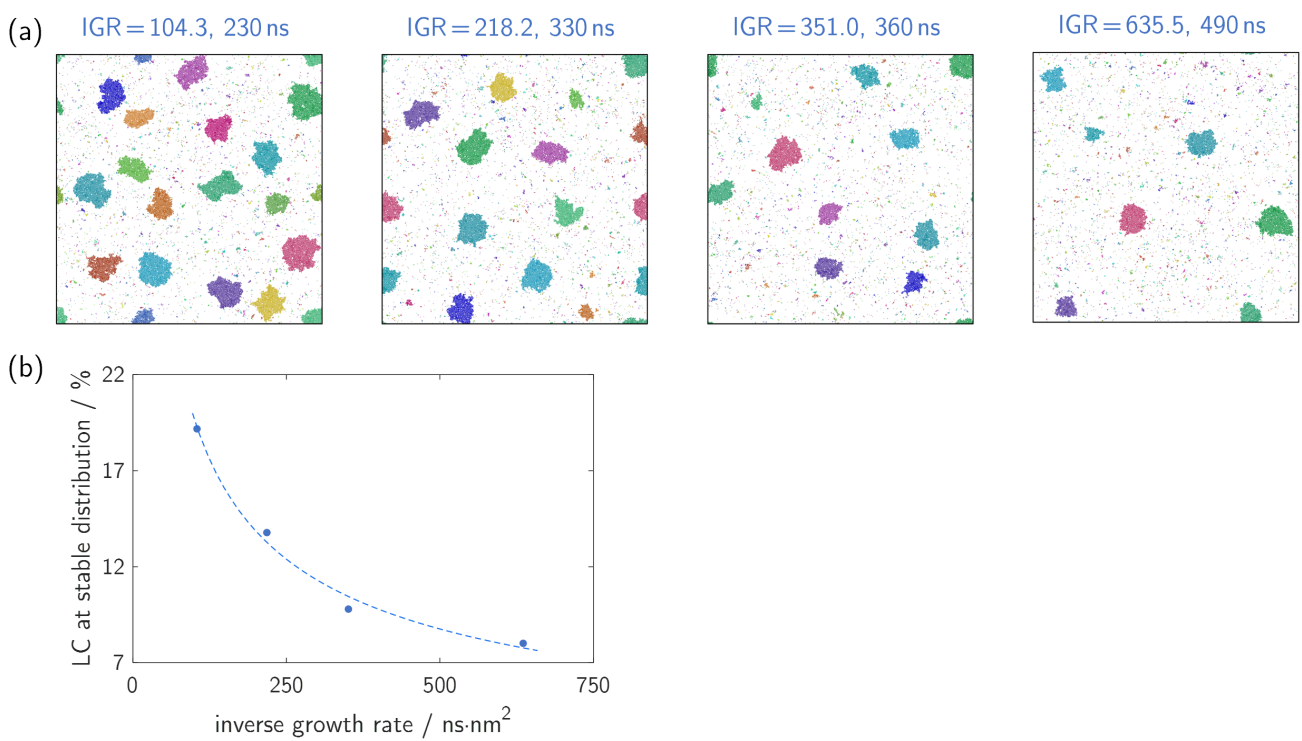

Figure S4. Molecules in layer 1 right after stable distribution is established. (a) horizontal distribution of molecules in layer 1 right after stable distribution established, simulated under four different IGR. The IGR and time corresponding to each image are labeled in the figure (the unit of IGR is ns·nm²). Molecules are colored by the clusters they belong to. Color choices of different images are unrelated. (b) The relation between CR (right after stable distribution established) and IGR. Four data points correspond to the four images in (a). The dashed curve serves as a guide to eyes, not a serious fitting.

# 6 CLC and its relation to region size

In the "growth mechanism of ice" section of main text we claim that high LC prefers cluster formation, and the critical LC (CLC) is a constant unrelated to region size in the thermodynamic limit. However, we don't provide direct evidence there. In this section, we will study the relation between CLC and region size directly by MD simulations.

To this end, we shall first slightly extend the definition of LC. In the main text the region LC defined on is the base cluster. However, it's hard to accurately control the LC on a cluster, and the cluster size continuously changes during ice growth, making quantitative study difficult. Instead, it's more feasible to study the relation between LC and region size in a full-bilayer model under NVT simulations. In such context, we can define LC as the following:

$$LC(M) = \frac{N_{L1}(M)}{A(M)\,\sigma_{perfect}}$$

where $M$ represents a full-bilayer model, $N_{L1}(M)$ is the number of molecules in layer 1 (red molecules in Fig. 1(a) of main text), and $\sigma_{perfect}$ is same as the one in the main text. The definition here follows the same concept as the one in the main text, except applying for full-bilayer models instead of clusters. Though the boundary of full bilayer models is different from that of clusters, such difference is negligible under thermodynamic limit where the region is very large.

We constructed multiple model systems beginning from perfect $I_h$ ice crystals same as the "QLL under equilibrium" section of main text, but with varied combinations of size and LC (size of 70, 140, 210, and 280 nm; LC varied from 2.9% to 4%, in layer 1). The LC of models are controlled by removing molecules from their surfaces, after an NVT relaxation of 3.75 ns. All models are constructed with no initial clusters to avoid statistical biases, and each with a reflection layer to avoid losing molecules. The simulations are done under NVT ensemble for 150 ns, with the LC and maximum cluster size of each model traced during the whole simulation.

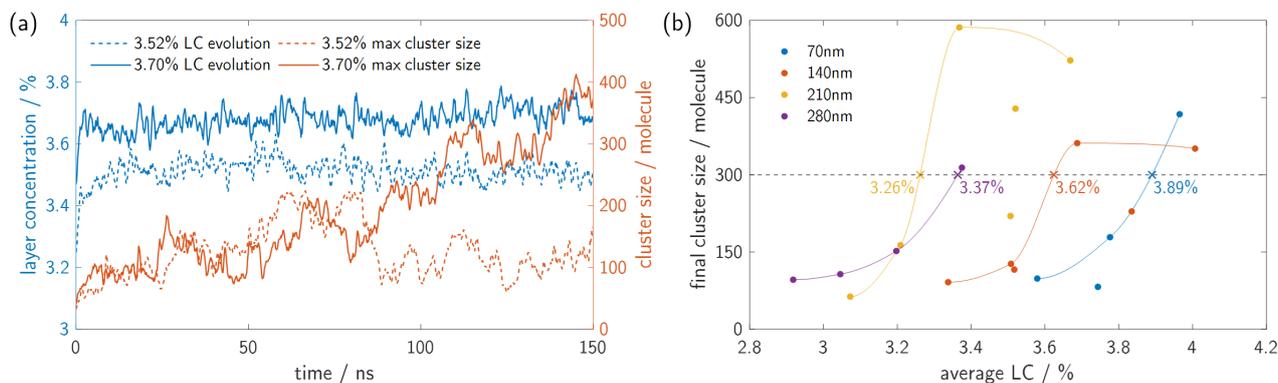

*Figure S5. Effect of LC and model size on cluster formation. (a) LC (blue) and maximum cluster size (red) of two typical simulations (average LC of 3.52% and 3.70%, both with the 140x140 nm model), as functions of time. See legends in the figure for details. The curves are slightly smoothed for better visual appearance. (b) relation between the final cluster size and average LC. Each independent simulation is represented by a dot, and its color represents the corresponding model size (see legends). Dots with each model size are connected by a curve with the corresponding color. The minimum LC required for forming stable clusters is estimated by the intersection of a curve and the horizontal dashed line (cluster size = 300). Intersections are shown by crosses in the figure, along with corresponding LC labeled. Note that some data points with "too low" final cluster sizes are not considered when connecting curves, because they show lower potential on forming clusters than simulations with lower LC, which is unlikely to be physical (we believe the potential of forming clusters should increase with LC).*

Fig. S5(a) shows the temporal evolution of LC and maximum cluster size in two typical simulations. The LC of both simulations keep steady most time, showing that our definition of LC is well-behaved. In the simulation with higher LC (3.70%, solid lines in Fig. S5(a)), the maximum cluster size increases consistently over time, implying that stable clusters can form under this LC (and model size). In the simulation with lower LC (3.52%, dashed lines in Fig. S5(a)), however, the maximum cluster size goes up only for a while and then goes down again, implying that current LC (and model size) cannot support forming stable clusters. Such observation indicates that high LC favors cluster formation.

Fig. S5(b) shows the results of all simulations in one figure. Each simulation is characterized by two values (along with model size):

(1) average LC, calculated over last 80% of the simulation;

(2) final cluster size, averaged over the maximum cluster size of the last 10% simulation. This is the measure of whether stable clusters may form.

Due to the stochastic nature of such simulations (technically "forming clusters" are somewhat rare events), some abnormal behaviors are observed in Fig. S5(b) (e.g. simulation with same model size and higher LC shows lower final cluster size). However, we can still tell that high LC and large system sizes generally favor the formation of clusters.

Now we shall turn to the relation between CLC and region size. In the main text we claim that theoretically the CLC should be a constant when region size reaches thermodynamic limit. More specifically, when the region is sufficiently large (so numerous molecules exist in it), forming a cluster in the layer would only have a marginal effect on the LC of other locations in the layer. That means forming clusters effectively becomes a local behavior (have no effect elsewhere), thus its feasibility only depends on local LC and is irrelevant to the region size. An intuitive analogy for this is the crystallization from solutions, whose feasibility only depends on the concentration of solution and is irrelevant to its volume (as long as it's large enough so crystallization itself has negligible effect on the concentration).

This conclusion is also supported by the MD simulation above. The CLC under each model size can be estimated from results in Fig. S5(b), if we ignore abnormal data points (see footnotes of Fig. S5). Fig. S5(b) shows the estimated CLC for all four model sizes, using 300 molecules as the criterion of forming stable clusters. We found that the CLC is negatively correlated with system size in 70~210 nm, showing that large models can form clusters under lower LC. However, the difference of CLC between 210 nm and 280 nm is small (and the order is "reversed": the *larger* 280 nm model has a *higher* estimated CLC). The reversed order is likely unphysical, but it also implies that the effect of system size on CLC would be small in large systems (so stochastic factors may dominate), thus supporting our conclusion. This also suggests that the thermodynamic limit is achieved when region size reaches ~210~280 nm, and suggesting a CLC of ~3.3%. In Section 2.3 of main text we used a slightly lower value (3.1~3.3%) for CLC under thermodynamic limit, since such simulations may fail to catch some cluster formation events thus tend to overestimate CLC.

# References

1. Bishop, C. L. et al. On thin ice: surface order and disorder during pre-melting. *Faraday Discuss.* **141**, 277-292 (2009).

2. Sánchez, M. A. et al. Experimental and theoretical evidence for bilayer-by-bilayer surface melting of crystalline ice. *Proc. Natl. Acad. Sci. U. S. A.* **114**, 227-232 (2017).

3. Pickering, I., Paleico, M., Sirkin, Y. A. P., Scherlis, D. A. & Factorovich, M. H. Grand Canonical Investigation of the Quasi Liquid Layer of Ice: Is It Liquid? *J. Phys. Chem. B.* **122**, 4880-4890 (2018).